\documentclass[aps,prl, reprint, showpacs,superscriptaddress,floatfix,amsmath,amssymb]{revtex4-1}

\usepackage{cancel}
\usepackage{pifont}
\usepackage{MnSymbol}
\usepackage{graphicx}
\usepackage{bm}

\usepackage{color}

\begin{document}

\title{Primordial Gravitational Waves in the Cosmic Bubble Chamber}
\author{Raymond Ang\'elil}\email{raymond@angelil.ch}
\address{Institute for Computational Science, University of Zurich, Winterthurerstrasse 190, 8057 Zurich, Switzerland}
\address{Deloitte AG, Risk Advisory, General-Guisan-Quai 38, 8022 Zurich, Switzerland}
\address{Credit Suisse AG, Uetlibergstrasse 231, 8070 Zurich, Switzerland}
\author{Prasenjit Saha}
\address{Physics Institute, University of Zurich, Winterthurerstrasse 190, 8057 Zurich, Switzerland}
\author{J\"urg Diemand}
\address{Physics Institute, University of Zurich, Winterthurerstrasse 190, 8057 Zurich, Switzerland}

\date{\today}

\begin{abstract}
We explore the effect of relic gravitational waves on the primordial phase change from the quark-gluon plasma into the low density hadron gas that occurred approximately $\sim 10^{-5}s$ after the beginning. We show that the gravitational wave, through doing work on the fluid, modulates the local volumes, causing a pressure modulation which either suppresses or boosts the bubble cavitation rate. The boosted rate is significant, implying that the phase transition could have occurred earlier than if this interaction is not considered. 
\end{abstract}

\keywords{Suggested keywords}

\maketitle

\section{Introduction}
In a stretched liquid, the phase transition from a homogeneous liquid into a vapor state is driven by the formation and subsequent growth of bubbles. In the early universe, the phase change from a primordial quark-gluon plasma into a low-density hadron gas occurred when the pressure, driven by gravitational expansion, dropped enough to turn the liquid into a superheated state\citep{earlyUniversePhaseTransitions,Mukhanov}. Driven by thermal fluctuations, the superheated liquid underwent a phase change into a gas by cavitation, producing bubbles larger than a critical size. The well-known gravitational effect which caused the drop in pressure is the radiation-domination expansion of the early universe: as the universe grew, the fluid density and therefore pressure dropped, precipitating the transition. A less-studied gravitational effect is the influence of primordial gravitational waves on the pressure: gravitational waves do work on the fluid by modulating the density through gravitational volume modulations. The short-lived, locally-expanded regions are in a low-pressure state, resulting in a superheated fluid region, which will cavitate into a gas by forming critical bubbles. We find that the nucleation rate in the presence of a gravitational wave $J_{GW}$, compared to a flat, Minkowskian background environment $J_{M}$, is boosted by a factor
\begin{equation}
\log{\frac{J_{GW}}{J_M}} \sim  \frac{\epsilon^2}{k},
\end{equation}
where $\epsilon$ is the amplitude of the wave, and $k$ the wavenumber. This means that the phase transition in the presence of gravitational waves, for most patches of the universe, happens significantly earlier than if the wave were not present. 

This letter explores the role of relic gravitational waves as drivers of early universe bubble formation, with the universe age approximately $t \sim 10^{-5}\,$s at an energy scale around $T\sim 300\,$MeV.
 We will do this by considering the effect of the wave on a homogeneous patch of primordial fluid, at a time in the expansion history when it has just entered the superheated state, yet one in which the nucleation rate is still low. We blanket the patch in a gravitational wave, and show that it boosts the nucleation rate higher than if no wave were present. To enable this calculation, we will proceed by making some assumptions on the equation of state of the fluid, derive the volume modulation effect of the gravitational wave, and then apply both to the formalism of classical bubble nucleation theory\citep{classicCav,tanaka}. 

In bubble chamber experiments, cavitation is used as a detection tool for high energy sub-atomic particles, such as dark matter searches\citep{xenon1T, bubbleDenzel}. However, there is much still to be learned about bubble cavitation itself. Bubble nucleation theory is built on the classical model of bubble cavitation\citep{Kalikmanov2013, Brennen, Kashchiev2000, BlanderKatz,kelton2010nucleation}.  Theoretical advances which predict nucleation rates and bubble growth rates have been tested in the lab in simple liquids\citep{bubbleLab, bubbleLab2, bubbleLab3}, and more recently in molecular dynamics simulations\citep{moreBubbleSim,tanaka,angelilBubbles,diemandBubbles, extended}. While the numerical experiments suggest minor modifications to the classical treatment, for our purposes, applying concepts of classical bubble cavitation to the primordial quark gluon plasma to hadron phase change, suffices in complexity for a first calculation.

The existence of gravitational waves is secure both from their indirect consequences, such as the slow orbital in-fall of binary pulsars \citep{gravWaveIndirect1, gravWaveIndirect2}, as well as from their direct detection \citep{gravWaveDirect} of binary black hole merger by ground based interferometers. 

However, the existence of primordial gravitational waves is not yet confirmed. Their discovery could confirm the inflationary model of cosmology - an early period of exponential expansion which is conjectured to have occurred sometime between $10^{-33}$s to $10^{-32}$s after the big bang. The inflationary epoch is a scenario inserted into the early cosmology time-line in order to solve a variety of problems which the non-inflationary cosmological model is unable to solve\citep{planckInflation}. The lack of direct evidence, as well as the imaginative requirements on the theoretical description of the inflatory dynamics at an energy scale far beyond those accessible in particle accelerators have naturally given rise to criticism of the inflationary scenario\citep{inflationCriticism}. The inflationary phase produces not just density inhomogeneities which form the seeds which eventually become the modern structure of the universe, but in doing so, produces gravitational waves. The discovery of primordial gravitational waves could provide indirect evidence for the inflationary paradigm. The search for primordial gravitational waves from their effects on the cosmic microwave background polarization is underway\citep{Bicep2, primWavesDetection}, and direct detection by space-based gravitational interferometers\citep{1975JETP...40..409G, 1979ZhPmR..30..719S,LisaPrimordialWaves} could be possible. 

\section{Pressure fluctuations from Gravitational waves}

Primordial gravitational waves arise from the quantum fluctuations of the gravitational field. Inflation stretches them out and makes their amplitudes even larger than the QCD energy scale. The waves are multi-modal, and travel in all directions. For the purposes of a primer calculation, we restrict ourselves to a single mode. A gravitational wave is a modulating perturbation of the space-time, with a metric
\begin{equation}\label{mywave}
g_{\mu\nu} = \eta_{\mu\nu} +\underbrace{ \epsilon \left( \begin{array}{cccc}
0 & 0 & 0 & 0  \\
0 & 1 & 0 & 0 \\
0 & 0 & -1 & 0 \\
0 & 0 & 0 & 0 \end{array} \right) \cos\left[k\left(z - t\right)\right]}_{h_{\mu\nu}},
\end{equation}
with a wavenumber $k$, and amplitude $\epsilon$. We have chosen one traveling in the $z$ direction, and we have also chosen the TT gauge to represent it. We have restricted ourselves to a time and spatial scale over which the expansion is not relevant i.e., one with 
\begin{equation}\label{eq:HubbleK}
k\gg \mathcal{H}.
\end{equation} We are concerned with the volume changes that the gravitational wave induces. These volume changes we will observe over some box-like region, with lengths $L_X$, $L_Y$, and $L_Z$. 

\begin{eqnarray}
V\left(t\right) &=& \int \sqrt{-g}d^3x\\
 &=& \int_0^{L_Y}\int_0^{L_X}\int_0^{L_Z}\left(1-\frac{\epsilon^2}{2}\cos\left(kz-\omega t\right) \right)dx dy dz \nonumber\\
 &=& L_XL_Y\left[L_Z - \frac{\epsilon^2}{2k}\sin\left(k\left(z-t\right) \right) \right] \nonumber
\end{eqnarray}
If we consider a spherical patch (although the shape is not important) approximately a quarter of the wavelength, we have extremal volume fluctuations 
\begin{equation}\label{eq:volfluc}
V = V_M \left(1\pm \frac{\epsilon^2}{2k}\right), 
\end{equation} 
where $V_M$ is the `Minkowski' volume - i.e., the volume in the absence of gravitational waves. 
\section{Gravitational-wave driven cavitation}
The effect of the gravitational wave is to modulate the volume occupied by the fluid. This results in a modulation of the pressure. The equation of state of the fluid is
\begin{equation} \label{eq:idealgas}
P \sim V^{-1}.
\end{equation} 
By \eqref{eq:volfluc}, the induced pressure fluctuation extrema are
\begin{equation} 
P = \frac{P_M}{\left(1\pm \frac{\epsilon^2}{2k}\right)}.
\end{equation}

Classical bubble nucleation theory assumes that if there is an energy barrier for the phase transition, we can see it in the Gibbs free energy of formation for a bubble of size $r$ 
\begin{equation}
\Delta G\left(r\right) = 4\pi r^2 \gamma_{QCD} - \frac{4\pi}{3}r^3 \left(P_{eq} - P\right),
\end{equation}
where $\gamma_{QCD}$ is the surface tension of the quark gluon - hadron gas interface, and $P_{eq}$ is the fluid's equilibrium pressure. Under the assumption of mechanical equilibrium at the critical size $r=r_*$, the corresponding bubble nucleation rate is proportional to the abundance of critical bubbles ($n_L$) times a kinetic factor\citep{tanaka, BlanderKatz,Brennen} \footnote{Refer to equation (12) in Blander \& Katz (1975)\citep{BlanderKatz}} 
\begin{equation}
J = n_L \left[\frac{2\gamma}{\pi m} \right]^{1/2}\exp{\left(-\frac{\Delta G_{r=r_*}}{kT}\right)},
\end{equation}
where $n_L$ is the number density.
Comparing the nucleation rate under the presence of gravitational waves to their absence, in extremal regions,
\begin{eqnarray}
\frac{J_{GW}}{J_M} &=& \exp{\left[-\Delta G_{M} - \Delta G_{GW} \right]}\\
&=& \exp\left[\pm P_M \frac{4\pi}{3} r_*^3 \frac{\epsilon^2}{2k} \right],
\end{eqnarray}
at a slightly modified critical size
\begin{equation}
r_{*, GW} = \frac{2\gamma_{QCD}}{P_{eq} - P_M\left(1 \pm \frac{\epsilon^2}{2k}\right)}.
\end{equation}
The plus-minus indicates the suppression or boost in the squeezed or stretched regions respectively. The critical bubble size is constrained by the size of the extremal wave fluctuation:
\begin{equation}
k^{-1} > r_{*}.  
\end{equation}
The effect of the gravitational wave as a driver of bubble nucleation is relevant only when the wavelength is long enough compared to the critical bubble size because bubbles formed within a volume extremum of a quarter wavelength will collapse. This puts a cutoff on the region of the primordial gravitational wave spectrum relevant to this process. From \eqref{eq:HubbleK}, compared to the Hubble scale, this process is sensitive to patches with 
\begin{equation}
r_{*} << \mathcal{H}^{-1}.
\end{equation}
\section{Discussion}
Laboratory-based bubble chamber experiments operate by detecting bubble nucleation events. The events occur due to localized energy deposits from high energy particles which interact with atomic nuclei, causing a local increase in temperature, allowing a bubble to form larger than the critical size. In accordance with classical nucleation theory, the sharp dependence that the nucleation rates have on thermodynamic variables, makes bubble chambers a powerful experimental tool\citep{bubbChamber1,bubbChamber2}.

Our apparatus is similar - yet instead of a confined chamber, and local temperature deposits acting as the thermodynamic driver behind bubble formation, our chamber is everywhere, and local volume deposits, from gravitational waves, drive the rates. In regions of increased volume, the energy barrier for bubble nucleation is lowered, increasing the nucleation rate, allowing the phase change to occur earlier, with a decreased critical bubble size.

For this process to be possible, the relic gravitational wave has done work on the fluid $W = -p \textrm{d}V$. This is fundamentally different to the way in which a gravitational wave interacts with a light-based interferometer. In an interferometer, unlike in a gravitational wave bubble chamber where the signal \eqref{eq:volfluc} is proportional to the first second power of the wave amplitude $\sim \epsilon^2$, the observed signal is proportional to the first power $\sim \epsilon$, and no work is done on the experimental apparatus\citep{angelilSaha}. Because the relic gravitational wave has done work on the fluid, the gravitational waves are attenuated, and that the fluid has been heated by the waves beyond the latent heat\footnote{The authors acknowledge that there are recent arguments that the phase transition is a cross-over and not a first order phase transition\citep{FLORtKOWSKI2011173, doi:10.1063/1.4875317, Schettler:2010wi}.} that is anyway associated with the phase transition in the absence of gravitational waves. This affects both super- and sub-critically sized bubbles. The primordial gravitational wave background must therefore decay as the energy is deposited. This could affect prospects for the direct detection of primordial gravitational waves with space-based interferometers, or their indirect detection through cosmic microwave background polarization modes.

The relevance of the effect explored in this letter depends on a number of poorly measured parameters. This includes both properties of the primordial gravitational radiation, as well as the thermodynamic quantities of the quark-gluon plasma fluid. We have considered only a single gravitational mode, yet the entire primordial gravitational wave spectrum $\epsilon_n$ and $k_n$ is relevant to the magnitude of the nucleation rate. 

The thermodynamic properties of the quark-gluon plasma are not well known, such as the planar surface tension $\gamma_{QCD}$, nor is whether or not a Tolman correction is relevant\citep{Tolman}. A further key parameter, is the equilibrium pressure $P_{eq}$ of the quark-gluon fluid, as this is temperature dependent, and this, as well as other fluid properties are affected by the quark-gluon wave fluid. The equation of state of the primordial quark gluon plasma is not yet well understood, although advances have has been made from heavy ion collisions, as well as progress in lattice QCD calculations\citep{hadronEoS}. More work is necessary to understand such a fluid and its equation of state.

In this letter we have looked at how relic gravitational waves work to drive bubble formation. A further mode of interaction with the fluid is via collapsing bubbles, or bubbles colliding into each other, and in doing so producing gravitational waves. At some point in the expansion history, there is therefore an equilibrium between the quark-gluon fluid and the gravitational waves: gravitational waves are not longer part of the background, instead, we have a short-lived quark-gluon-gravitational wave fluid, with its own thermodynamic properties. As the universe grew and cooled, they decouple from one another and the phase change proceeds. It is unclear to the authors to what extent this scenario affects the phase transition.

\bibliographystyle{apsrev4-1}
\bibliography{gravitational_wave_cavitation}

\end{document}